\documentclass[pra,preprint]{revtex4-1}
\usepackage{graphicx}
\usepackage{amsmath}
\usepackage{bm}
\usepackage{array}
\usepackage{braket}
\usepackage{microtype}
\usepackage{wasysym}
\renewcommand{\baselinestretch}{1.0}

\begin{document}

\renewcommand{\baselinestretch}{1.0}

\title{Time-Correlation and Decoherence in a Two-Particle Interferometer}

\author{Bert\'{u}lio de Lima Bernardo}

\affiliation{Departamento de F\'{\i}sica, CCEN, Universidade Federal 
da Para\'{\i}ba, Caixa Postal 5008, 58059-900, Jo\~ao Pessoa, PB, Brazil}

\email{bertulio@fisica.ufpb.br}

\begin{abstract}
  A two-particle interferometer is theoretically analyzed, to show how
  decoherence induced by interactions with the environment affects
  time correlations, a process we call time-correlation decoherence.
  Specifically, on the basis of simple mathematical analysis we show
  how the interaction between a bipartite entangled system and a
  photon bath representing the environment can efface the oscillations
  in the coincidence-detection rate of the interferometer. We discuss
  the dependence of this kind of decoherence on the photon energy
  and density.
\end{abstract}


\maketitle


\section {Introduction}
\label{sec:1}
As famously stated by Feynman \cite{feynman}, Young's double-slit
experiment has in it the {\it heart of quantum mechanics} and {\it
  contains the only mystery of the theory}. From the
quantum-mechanical point of view, this experiment consists of a group
of particles, such as electrons, approaching a screen with two
slits. After traversing the slits, the particles impinge on a distant
detector screen, which registers permanently their positions. If no
information is available concerning the passage of the particles
through the slits, the particle density on the detection screen
displays an interference pattern described by the expression
$\rho(x)=\frac{1}{2}|\psi_{1}(x)+\psi_{2}(x)|^{2}$, where
$\psi_{1}(x)$ and $\psi_{2}(x)$ are the partial wave functions
associated with the passage through slits 1 and 2, respectively. On
the other hand, if the experimental procedure determines the slit
traversed by each particle, the interference pattern disappears and
the detector exhibits the classical addition of two patterns, one due
to the particles that have traversed slit 1 and the other due to those
that have traversed slit 2, i.~e.,
$\rho(x)=\frac{1}{2}|\psi_{1}(x)|^2+\frac{1}{2}|\psi_{2}(x)|^{2}$.
This experiment leads to the conclusion that quantum interference
is incompatible with which-path information.
        
Let us now consider the behavior of a classical macroscopic object
immersed in a large environment of gaseous molecules, light, thermal
photons, etc.. At any moment a huge number of environmental particles
collide with the object, in such a way that they will carry some
information about the object, on its position and orientation in
space, for instance. In this case, the information is associated with the
scattering positions and deflection angles. We see that every object
interacts with its environment, as a result of which information about
the physical properties of the former is inevitably encoded in the latter.

Interactions between quantum objects and their environments are
significantly weaker, because quantum systems are several orders of
magnitude smaller than classical ones. Nonetheless, system-environment
interactions are ubiquitous in quantum physics and can transfer
which-path (or which-state) information to the environment by the
aforementioned mechanism. In other words, an interacting environment
suppresses interference (wave-like behavior) in atomic systems and
consequently bars quantum manifestations at the macroscopic
scale. System-environment interactions explain how the classical
behavior of the macroscopic world emerges from the quantum properties
of its building blocks \cite{schloss, zurek, schloss2}.

A quantum superposition depends on the relative phases between its
components. System-environment interactions transfer which-path (or
which-state) information to the environment at the expense of the
coherence among those relative phases. This inevitable monitoring of
the system by the environment therefore amounts to the so-called
\emph{decoherence} process. In the two-slit interferometer, one can
always regard any mechanism offering information on the particle path
as a form of system-environment interaction responsible for a specific
kind of decoherence, i.~e., the destruction of the interference pattern on
the detection screen \cite{wootters, schloss2}. This remarkably
simple, evident form of decoherence is the object of our analysis.

We are particularly interested in a class of interference devices
first developed in the 1980's, the two-particle
  interferometer \cite{rarity,kwiat,ou}. Certain experiments have
shown that when two entangled particles separately go through a
single-particle interferometer, such as the Young interferometer,
an interference pattern results when the rate of coincident arrival is
measured, while no such pattern appears when only one particle is
observed.\cite{mandel} Entangled particles, in this context, are
particles simultaneously created by the same source in such a way
that, due to momentum conservation, one only has to determine the
position of one particle to predict the position of the other. This
kind of interferometer, as we shall see, is particularly sensitive
to entanglement correlation. Nonetheless, the literature contains no
detailed description of decoherence in two- and many-particle systems of this kind.

The purpose of this paper is to establish a simple, direct connection
between decoherence and two-particle interferometry.  To this end, we
discuss a \emph{gedanken} experiment originally devised by Horne and
Zeilinger \cite{horne}, which is convenient because simple
calculations suffice to describe the system after interaction with the
bath of monochromatic photons that here represents the environment. As
already mentioned, in the two-particle interferometer under study
interference is only observed in time-correlation measurements.  For
this reason, we will refer to the environmental disturbance as
\emph{time-correlation decoherence} (TCD), to distinguish it from the
well-known spatial decoherence that is commonplace in the
single-particle systems.

\section{Two-particle interferometry}
\label{sec:2}
The \emph{gedanken} experiment, which Gottfried has also explored very
well \cite{gott}, analyzes the particles produced by the decay process
$A \rightarrow a + b$, each daughter particle going through a
double-slit apparatus, as shown in Fig.~1. If $A$ is approximately at
rest, momentum conservation forces $a$ and $b$ to travel in
approximately opposite directions. Therefore, if $a$ passes through
one of the slits on the right, $b$ must pass through the 
diametrically opposite slit on the left.

\begin{figure}[htb]
\begin{center}
\includegraphics[height=1.6in]{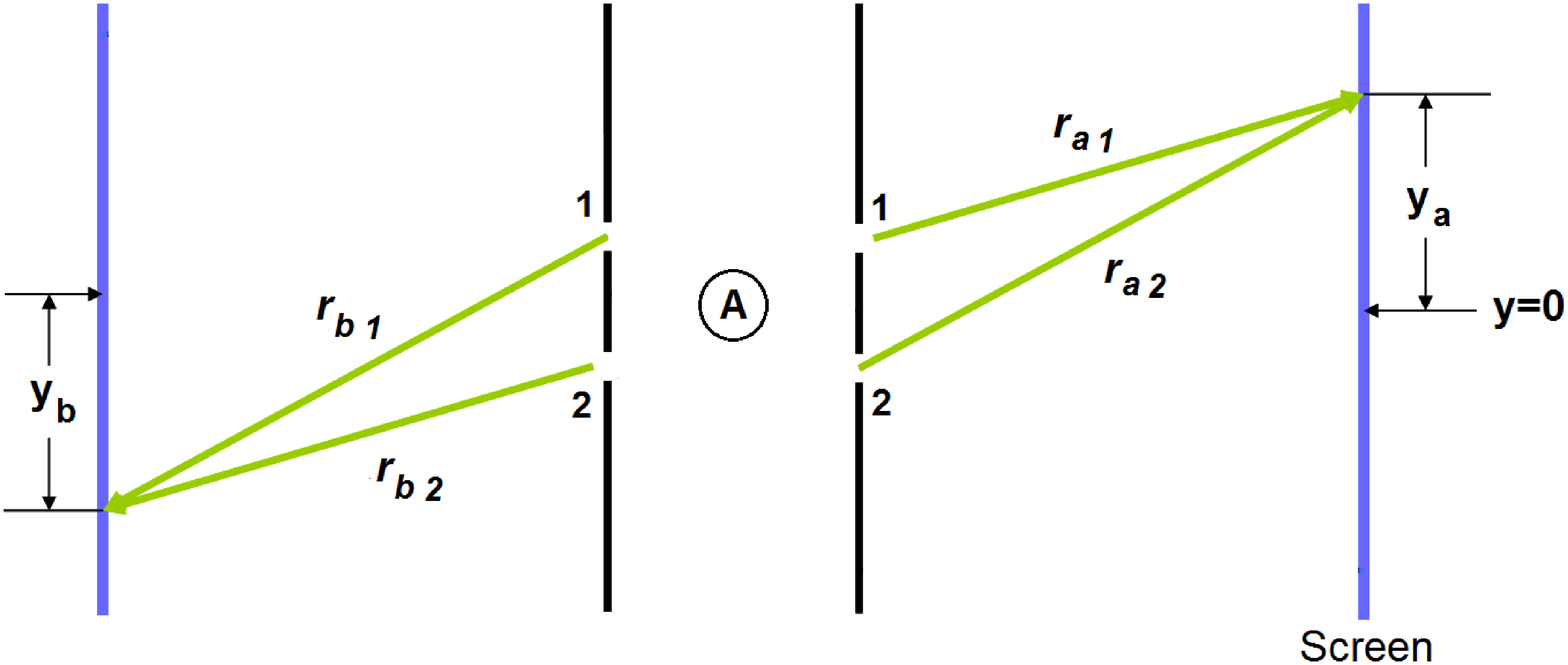}
\end{center}
\label{F1}
\caption{(Colored online) Schematic representation of the
  \emph{gedanken} experiment discussed in the text. Particle $A$, with
  approximately zero momentum, decays into two particles $a$ and
  $b$. To conserve momentum, the daughter particles must travel in
  approximately diametrically opposite paths. Each particle traverses
  a double-slit apparatus before being detected by a screen, at
  $y_a$ or $y_{b}$. The lengths $r_{a1}$, $r_{a2}$, $r_{b1}$, and
  $r_{b2}$ are the distances from the slits to the detection points.} 
\end{figure}

Let $\ket{R_{1}}$ and $\ket{R_{2}}$ denote the quantum states of
particle $a$ corresponding to passage through slits 1 and 2 on the
right, and $\ket{L_{1}}$ and $\ket{L_{2}}$ denote the quantum states
of particle $b$ corresponding to the passage through slits 1 and 2 on
the left, respectively. We can then write the two-particle entangled quantum state
in the form
\begin{equation}
\label{1}
\ket{\psi}= \frac{1}{\sqrt{2}}(\ket{R_{1}}\ket{L_{2}} + \ket{R_{2}}\ket{L_{1}}). 
\end{equation}

On the right-hand side of Eq.~\eqref{1} we recognize a state that is
entangled, in the above-defined sense, since $\ket{\psi}$ cannot be
factorized into a simple product of $a$ and $b$ states, i.~e., no two
states $\ket{R_{i}}$ and $\ket{L_{j}}$ can be found such that
$\ket{\psi} = \ket{R_{i}} \ket{L_{j}}$. The state of one particle
cannot be specified without reference to the other particle; the two
particles are therefore entangled.

The concepts of density matrix and reduced density matrix have
capital importance in decoherence theory. We therefore adopt those
concepts from the outset, to familiarize the reader with them. We
shall make only simple use of these tools. To compare our formalism
with the quantum-state formalism, we recommend Gottfried's analysis of
the same \emph{gedanken} experiment \cite{gott}. A clear
introduction to the density matrix and reduced density matrix in the
context of decoherence can be found in Ref.~[12].

To start, we write the density matrix $\rho = \ket{\psi}\bra{\psi}$
for this system in the following form
\begin{equation}
\label{2}
\rho= \frac{1}{2}\sum_{\substack{ij=1\\i \neq 
j}}^{2}\ket{R_{i}}\ket{L_{j}}\bra{L_{j}}\bra{R_{i}} 
+ \frac{1}{2}\sum_{\substack{ij=1\\i \neq
j}}^{2}\ket{R_{i}}\ket{L_{j}}\bra{L_{i}}\bra{R_{j}}. 
\end{equation}

To describe the behavior of one of the particles, the reduced density
matrix associated with that particle is convenient. If we are
interested in particle $a$, for instance, to compute the reduced
density matrix $\rho_{a}$ we trace Eq.~\eqref{2} over
the states of particle $b$ in the following way:
\begin{equation}
\label{3}
\rho_{a} = \operatorname{Tr_{b}}\ket{\psi}\bra{\psi} 
= \frac{1}{2}\sum_{i=1}^{2}\braket{L_{i}|\psi}\braket{\psi|L_{i}}.   
\end{equation}

Since $\braket{L_{1}|L_{2}}=0$, we have that $\rho_{a}=
1/2\sum_{i=1}^{2}\ket{R_{i}}\bra{R_{i}}$. This density matrix
corresponds to a particle density $\rho(y_{a})$ on the detecting
screen on the right-hand side of Fig.~1 given by the expression
\begin{equation}
\label{4}
\rho(y_{a}) \equiv \braket{y_{a}|\rho_{a}|y_{a}}
= \frac{1}{2}|\psi_{a1}(y_{a})|^{2} + \frac{1}{2}|\psi_{a2}(y_{a})|^{2},    
\end{equation}
where $\psi_{ai}(y_{a})= \braket{y_{a}|R_{i}}$ ($i=1,2$). 

As we can see, the distribution of $a$ particles on the detection
screen exhibits no interference pattern.  Given the symmetry of the
apparatus, we see that \emph{mutatis mutandis} the same result
describes the distribution of $b$ particles on the left-hand detection
screen. Physically speaking, the absence of interference patterns
stems from assuming particle $A$ to be approximately at rest
initially. According to the uncertainty principle, we have almost no
information on the initial position of $A$.  Particle $A$ is therefore
equivalent to a large source of daughter particles, and the path of
each daughter is undefined relative to the path of the
other. Consequently, single-particle interference cannot occur.

Let us now analyze the system as a whole. The probability density of
simultaneously detecting particle $a$ at $y_{a}$ and particle $b$ at
$y_{b}$ is given by the expression
\begin{equation}
\label{5}
\rho(y_{a},y_{b}) \equiv \bra{y_{b}}\braket{y_{a}|\rho|y_{a}}\ket{y_{b}}.    
\end{equation}

Substitution of Eq.~\eqref{2} into Eq.~\eqref{5} yields the result
\begin{align}
\label{6}
\rho(y_{a},y_{b}) &= \frac{1}{2} 
\sum_{\substack{ij=1\\i \neq j}}^{2}
\braket{y_{a}|R_{i}}\braket{y_{b}|L_{j}}\braket{L_{j}|y_{b}}
\braket{R_{i}|y_{a}}\nonumber\\
&+\frac{1}{2}\sum_{\substack{ij=1\\i \neq j}}^{2}
\braket{y_{a}|R_{i}}\braket{y_{b}|L_{j}}\braket{L_{i}|y_{b}}
\braket{R_{j}|y_{a}}.    
\end{align}

Let us assume that, after passing through one of the slits, the
wavefunctions of the particles are spherical waves, i.~e., given by
the expressions
\begin{equation}
\label{7}
\psi_{aj}(r_{aj})=\braket{r_{aj}|R_{j}}=\frac{e^{ikr_{aj}}}{r_{aj}}    
\end{equation}
and
\begin{equation}
\label{8}
\psi_{bj}(r_{bj})=\braket{r_{bj}|L_{j}}=\frac{e^{ikr_{bj}}}{r_{bj}}\qquad(j=1,2),
\end{equation}
where the $r_{(a,b)j}$ denote the distances from the slits to the
detection points, and $k$ is the wavenumber.

If we let the distance between the slits and the detection screen be
much larger than the separation between the two slits so that we are
in the Fraunhofer diffraction limit, we have that $r_{a (1,2)} \approx L \mp
\theta y_{a}$ and $r_{b (1,2)} \approx L \mp \theta y_{b}$ \cite{born},
with the coordinates $y$ defined in Fig.~1, and the angle $\theta$ and distance $L$
defined in Fig.~2. Equations~\eqref{7}~and \eqref{8} then yield the
approximate equalities
\begin{equation}
\label{9}
\braket{y_{a}|R_{1,2}} \approx \frac{e^{ik( L \mp \theta y_{a})}}{ L \mp \theta
y_{a}}    
\end{equation}
and
\begin{equation}
\label{10}
\braket{y_{b}|L_{1,2}} \approx \frac{e^{ik( L \mp \theta y_{b})}}{ L \mp \theta
y_{b}}.   
\end{equation}

We now substitute Eqs.~\eqref{9} and~\eqref{10} into Eq.~\eqref{6}.
Notice taken that the denominators can all be absorbed into an
irrelevant overall factor, for small diffraction angles we can write
the following expression for the joint probability of detecting particle $a$ at
$y_{a}$ and
particle $b$ at $y_{b}$:
\begin{equation}
\label{11}
\rho(y_{a},y_{b}) \doteq \cos^{2}\big(k \theta(y_{a}-y_{b})\big),    
\end{equation}
where the symbol $\doteq$ stands for equality up to a constant
factor. This equation shows that the coincident-arrival rate is a
periodic function of the relative position $y_{a}-y_{b}$, a functional
form characteristic of interference. It is not difficult to identify
the source of this curious behavior: as Eq.~\eqref{1} shows, the
daughter particles emitted in opposite directions by the decay of
particle $A$ can reach the screens in two alternative ways.
The interference between these two paths is responsible
for the oscillatory coincidence rate.

The contrast between Eqs.~\eqref{4}~and \eqref{11} constitutes a
particular instance of a relation between the single-
and two-particle interferences first identified by Jaeger \emph{et
  al.}~\cite{jaeger} in their analysis of the system depicted in Fig.~1.
Ref.~\onlinecite{jaeger} showed that if the initial state is maximally entangled, the
coincidence rate is affected by interference, as we have shown, while
single-particle detection patterns are not.  By contrast, when the
particles are created in a separable state, only single-particle
interference arises. More specifically, the more entangled the initial
state, the stronger the interference in the coincidence detection
rate, and the weaker the single-particle interference. This
complementarity has been verified in a number of experiments \cite{rarity,
  kwiat, ou}.

Although capturing essential features of experimental systems,
Fig.~1 is schematic. One of its most important limitations is the
absence of interaction between the particles and the environment.  In
the next section we show, for the first time, how this interaction
undercuts the interference responsible for the oscillatory behavior in
Eq.~\eqref{11}, i.~e., how the system-environment interaction
gives rise to decoherence.

\begin{figure}[htb]
\begin{center}
\includegraphics[height=1.5in]{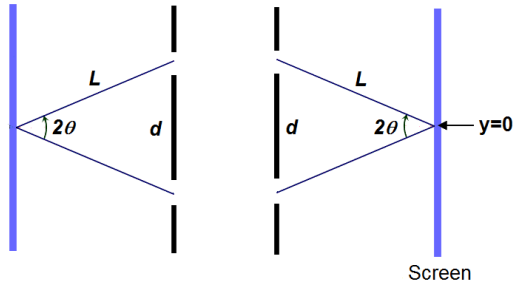}
\end{center}
\label{F2}
\caption{(Colored online) Distance $L$ and the angle $\theta$ associated with
  arrangement in Fig.~1.}
\end{figure}

\section{Time-Correlation Decoherence}

We now turn to the open system, i.~e., to particles that interact with
the environment. As we shall see, decoherence will arise, i.~e.,
time correlations will lose coherence.

First, we place a light source beyond the two slits on the right-hand
side of Fig.~1, to play the role of the environment. In this
arrangement, if we place a detector behind each slit, a photon that
happens to be scattered by particle $a$ near one of the slits will be
recorded by the nearest detector. The collision between particle $a$
and the photon represents the system-environment interaction, which,
as we shall show, entangles them and causes decoherence.  Next, we
restrict our analysis to photons with an wavelength so short that
diffraction can be ruled out and we can be sure that a photon
scattered at slit 1 cannot reach the detector at slit 2 or vice
versa. The purpose of the detectors is to detect interactions between
particle $a$ and the environment. In no way do they affect decoherence.

In analogy with the discussion in Section~\ref{sec:2}, we describe the
system-environment interaction by the expressions
\begin{equation}
\label{12}
\ket{R_{1}}\ket{L_{2}}\ket{\epsilon_{0}} \rightarrow
\ket{R_{1}}\ket{L_{2}}\ket{\epsilon_{1}}  
\end{equation}
and
\begin{equation}
\label{13}
\ket{R_{2}}\ket{L_{1}}\ket{\epsilon_{0}} \rightarrow
\ket{R_{2}}\ket{L_{1}}\ket{\epsilon_{2}},  
\end{equation}
where the environmental states $\ket{\epsilon_{0}}$,
$\ket{\epsilon_{1}}$, and $\ket{\epsilon_{2}}$ are the initial
photonic quantum state, the state into which the arrival of a photon
scattered near slit 1 triggers detector 1, and the state into which
the arrival of a photon scattered near slit 2 triggers detector 2,
respectively.  The initial environmental state $\ket{\epsilon_{0}}$
evolves into $\ket{\epsilon_{1}}$ or $\ket{\epsilon_{2}}$, depending
on the system state. Equations.~\eqref{12}~and \eqref{13} are valid
only if particle $a$ scatters a photon right after passing through one
of the slits, prior to any other collision. Otherwise, it would be
incorrect to write $\ket{R_{1}}$ or $\ket{R_{2}}$ (which, according to
Eq.~\eqref{9}, represent spherical waves emerging from slits 1 and 2)
on the right-hand sides of expressions~\eqref{12}~or \eqref{13},
respectively. The linearity of the Schr\"{o}dinger equation implies
the von Neumann measurement scheme \cite{zurek, schloss}
\begin{align}
\label{14}
\frac{1}{\sqrt{2}}(\ket{R_{1}}\ket{L_{2}} +
\ket{R_{2}}\ket{L_{1}})\ket{\epsilon_{0}} &\rightarrow
\nonumber\\ 
\ket{\phi}=\frac{1}{\sqrt{2}}(\ket{R_{1}}\ket{L_{2}}\ket{\epsilon_{1}}
&+  \ket{R_{2}}\ket{L_{1}}\ket{\epsilon_{2}}). 
\end{align}

We see that the system states have become entangled with the
environmental states, which encode information on the particle
paths. The initial coherence between the system states
$\ket{R_{2}}\ket{L_{1}}$ and $\ket{R_{1}}\ket{L_{2}}$ is now shared
with the environment, i.~e., is now a property of the system-environment state.

Let us analyze the behavior of this system in more detail. If we
determine the reduced density matrix $\rho_{a} =
\operatorname{Tr_{bE}} \ket{\phi}\bra{\phi}$ for particle $a$, where
$\operatorname{Tr_{bE}}$ stands for the trace over the states of
particle $b$ and the environment, and proceed to calculating
$\braket{y_{a}|\rho_{a}|y_{a}}$, it results that the probability
density $\rho({y_{a}})$ of finding particle $a$ at position $y_{a}$ on
the screen is still given by Eq.~\eqref{4}. Not surprisingly, the
entanglement between system and environment has no effect upon the
already-incoherent single-particle probability density.

On the other hand, if we calculate the reduced density matrix
$\rho_{ab} = \operatorname{Tr_{E}} \ket{\phi}\bra{\phi}$
 for the two particles, where $\operatorname{Tr_{E}}$
stands for the trace over the environmental states only, we find the equality
\begin{equation}
\label{15}
\rho_{ab} = \frac{1}{2}\sum_{k=1}^{2} \bra{\epsilon_{k}}O_{ij} +
Q_{ij}\ket{\epsilon_{k}},
\end{equation}
where
\begin{equation}
O_{ij}= \sum_{\substack{ij=1,\\i \neq j}}^{2}
\ket{R_{i}}\ket{L_{j}}\ket{\epsilon_{i}}\bra{\epsilon_{i}}\bra{L_{j}}\bra{R_{i}}\label{eq:1}
\end{equation}
and 
\begin{equation}
Q_{ij} = \sum_{\substack{ij=1,\\i \neq j}}^{2}
\ket{R_{i}}\ket{L_{j}}\ket{\epsilon_{i}}\bra{\epsilon_{j}}\bra{L_{i}}\bra{R_{j}}.\label{eq:2}
\end{equation}

If a given photon is recorded by detector 1, the same photon cannot be
recorded by detector 2. Mathematically, this self-evident notion
corresponds to the equality $\braket{\epsilon_{i}|\epsilon_{j}}=0$ for
$i \neq j$. Equation~\eqref{15} therefore reduces to the equation
\begin{equation}
\label{16}
\rho_{ab}=\frac{1}{2}\sum_{\substack{ij=1,\\i \neq
j}}^{2}\ket{R_{i}}\ket{L_{j}}\bra{L_{j}}\bra{R_{i}},  
\end{equation}
and Eqs.~\eqref{9}~and \eqref{10} yield the following expression for
the probability density of simultaneously detecting particle $a$ at $y_{a}$ and
particle $b$ at $y_{b}$:
\begin{equation}
\label{17}
\rho(y_{a},y_{b}) \equiv \bra{y_{a}} \braket{y_{b}|\rho_{ab}|y_{b}}\ket{y_{a}} =
\mathrm{const}. 
\end{equation}

The probability distribution in Eq.~\eqref{17} is position
independent. We therefore see that the system-environment interaction
has destroyed the coincidence-rate interference expressed by
Eq.~\eqref{11}, i.~e., the interference prevalent in the isolated,
photon-free system. Since time-correlation interference is lost, we
call this phenomenon time-correlation decoherence (TCD).

Interesting issues emerge when we examine the environmental
properties. In particular, we are insterested on the dependence of TCD
upon the photon energy, or equivalently, upon the wavelength of the
light. Up to this point we have only considered the small-wavelength
limit, i.~e., a wavelength $\lambda$ that is dwarved by the slit separation
$d$. With larger wavelengths, diffraction allows photons scattered
near slit 1 (2) to reach detector 2 (1). The light is now unable to
resolve the separation between the slits and the environment and cannot encode
a significant amount of information on the particle paths.

In order to account for these new possibilities, we now write the
system state of the system in the form
\begin{align}
\label{18}
\ket{\varphi} &= n \ket{R_{1}}\ket{L_{2}}\ket{\epsilon_{1}} 
+ m \ket{R_{1}}\ket{L_{2}}\ket{\epsilon_{2}}  \nonumber \\  
&+ n \ket{R_{2}}\ket{L_{1}}\ket{\epsilon_{2}} 
+ m \ket{R_{2}}\ket{L_{1}}\ket{\epsilon_{1}},  
\end{align}
where $n$ and $m$ are the probability amplitudes for a photon
scattered near a given slit to be recorded by the detectors that are
closer and farther from that slit, respectively. The right-hand side
of Eq.~\eqref{18} remains invariant under the change
$1\leftrightarrow2$ because we are working with identical slits and
symmetrically positioned detectors.

When we calculate the reduced density matrix,
$\rho_{ab}^{(\varphi)} = \operatorname{Tr_{E}}\ket{\varphi}\bra{\varphi}$, under the
condition $\braket{\epsilon_{1}|\epsilon_{2}} = 0$, the following
result emerges:
\begin{align}
\label{19}
\rho_{ab}^{(\varphi)} &= (|n|^2 + |m|^2) \sum_{\substack{ij=1,\\i \neq
j}}^{2}\ket{R_{i}}\ket{L_{j}}\bra{L_{j}}\bra{R_{i}}  \nonumber \\ 
&+ (nm^{*} + n^{*}m) \sum_{\substack{ij=1,\\i \neq
j}}^{2}\ket{R_{i}}\ket{L_{j}}\bra{L_{i}}\bra{R_{j}}. 
\end{align}

Equation~\eqref{19} is our central result. To find the probability
density of simultaneously detecting particle $a$ at $y_{a}$ and
particle $b$ at $y_{b}$ as a function of the amplitudes $n$ and $m$ we
only have to compute
$\bra{y_{b}}\braket{y_{a}|\rho_{ab}^{(\varphi)}|y_{a}}\ket{y_{b}}$. The
second term on the right-hand side of Eq.~\eqref{19}, which contains
the off-diagonal elements of $\rho_{ab}^{(\varphi)}$ on the basis $\{
\ket{R}\ket{L} \}$, is usually referred to as the interference term
because it monitors the quantum coherence among the components on the
right-hand side of Eq.~\eqref{18} \cite{schloss2}.

In the small-wavelength limit, $n=1/\sqrt{2}$ and $m=0$. In this case,
as expected, Eq.~\eqref{19} reduces to Eq.~\eqref{16}. TCD is
maximum and the coincidence arrival rate displays no vestige of
interference. In the large-wavelength limit, on the other hand, $n = m = 1/2$, since
the amplitude of a photon reaching a detector is independent of the slit
at which it was scattered. Under these conditions, Eq.~\eqref{19}
reduces to Eq.~\eqref{2}, TCD is minimum, and the coincidence arrival
rate shows the interference features identified in our discussion of
Eq.~\eqref{11}. No information on particle paths is conveyed to the
environnment. 

In the intermediate case, in which the detectors receive only partial
which-path information, we have that $n > m \neq 0$. The
coincidence-rate interference is weaker than in the large-wavelength
limit and the probability distribution combines a term 
reminiscent of Eq.~\eqref{11} with a constant contribution, analogous to
Eq.~\eqref{17}:
\begin{align}
\label{20}
\rho(y_{a},y_{b}) &\equiv \bra{y_{a}}
\braket{y_{b}|\rho_{ab}|y_{b}}\ket{y_{a}}  \nonumber\\ 
&\doteq |n|^{2} + |m|^{2} + (nm^{*} + mn^{*}) \cos\big(2k \theta (y_{a} -
y_{b})\big). 
\end{align}

Another important parameter is the intensity of the light source,
i.~e., the photon density in the region beyond the slits on the
right-hand side of Fig.~\ref{F1}. So far we have implicitly assumed
the intensity to be sufficiently high to insure scattering, with
100\,\% certainty.  This constraint relaxed, the properties of the
system are described by a mixed density matrix \cite{schloss2} of the
form $\rho = w_{1}\ket{\phi}\bra{\phi} +
w_{2}\ket{\alpha}\bra{\alpha}$. Here $\ket{\phi}$ is defined as in
Eq.~\eqref{14}, $\ket{\alpha} = \ket{\psi}\ket{\epsilon _{0}}$ is a
separable (non-entangled) system-environment state associated with the
absence of collisions, and $w_{1}$ and $w_{2}=1-w_{1}$ are the
classical probabilities of particle $a$ scattering or not scattering a
photon after passing through the slits, respectively. We calculate the
reduced density matrix, $\rho_{ab} = \operatorname{Tr_{E}}(\rho)$ and
from $\rho_{ab}$, with the wavefunctions in Eqs.~\eqref{9}
and~\eqref{10}, we find the following expression for the probability
density $\rho(y_{a},y_{b})$ to detect the two particles in
coincidence:
\begin{align}
\label{21}
\rho(y_{a},y_{b}) &\equiv \bra{y_{a}}
\braket{y_{b}|\rho_{ab}|y_{b}}\ket{y_{a}}  \nonumber\\ 
&\doteq w_{1} + 2 w_{2} \cos^{2}\big(k \theta(y_{a}-y_{b})\big), 
\end{align}
which, as expected, combines features found in
Eqs.~\eqref{11}~and \eqref{17}. 

For completeness, we cursorily discuss an alternative arrangement, with an
additional light source and two other detectors beyond the slitted
screen on the left-hand side of in Fig.~1. Qualitatively, the new
arrangement is equivalent to the setup we have analyzed.  As
long as one of the particles or both of them scatter photons after
passing through the slits, the environmental state changes as it
acquire information on the paths followed by the particles.  Clearly,
with two collision alternatives, the odds in favor of acquiring
which-path information are higher, and interference is weakened.
Equations~\eqref{17}, \eqref{18}, and \eqref{19} are still applicable,
but given that each particle can now collide with a photon, the
collision-probability parameter $w_{1}$ is larger. For example, if the two light
sources are identical $w_{1}$ is twice larger than in the previous case.

\section{Conclusion}

In conclusion, we have quantitatively studied a class of
quantum-mechanical decoherence processes, to show how the
system-environment interactions suppress coincidence-rate interference
in a two-particle interferometer. The environment was modeled by a
photon bath.  Given the loss in particle time-correlation coherence,
we have called this process time-correlation decoherence (TCD). In addition, we
have brought to light the decisive importance of the photon energy and
density in TCD.

\begin{acknowledgments}
  The author gratefully acknowledges Eric J. Heller for invaluable
  comments and discussions. This work was supported by
  Coordena\c{c}\~{a}o de Aperfei\c{c}oamento de Pessoal de N\'{i}vel
  Superior (CAPES) and by Conselho Nacional de Desenvolvimento
  Científico e Tecnológico (CNPq).
\end{acknowledgments}


\end{document}